\documentclass[12pt,a4paper,oneside]{article}
\usepackage[T1]{fontenc}                                                        
\usepackage{graphics} 
\usepackage{amsfonts}                                                           
\usepackage{palatino}                                                           
\usepackage{url}                                                           
\usepackage[english]{babel}                                                     
\usepackage{makeidx}
\newtheorem{theorem}{Theorem}
\newcommand{\teo}[2]{\begin{theorem}#1\ \\ {\bf Proof:} {\em #2}
\hfill $\Box$ \end{theorem}}

\title{Pathwords: a user-friendly schema for common passwords management} 
                                                                                
\author{Michele Finelli\thanks{{\sc BioDec} --- {{\tt m@biodec.com}}}}

\begin{document}

\maketitle
                                                                                
\date{}                                                                         
                                                                                
\abstract

Many computer-based authentication schemata are based on {\em
passwords}. Logging on a computer, reading email, accessing content on a
web server are all examples of applications where the identification of
the user is usually accomplished matching the data provided by the user 
with data known by the application.

Such a widespread approach relies on some assumptions, whose
satisfaction is of foremost importance to guarantee the robustness of
the solution. Some of these assumptions, like having a ``secure''
channel to transmit data, or having sound algorithms to check the
correctness of the data, are not addressed by this paper. We will focus
on two simple issues: the problem of using adequate passwords and the
problem of managing passwords.

The proposed solution, the {\em pathword}, is a method that guarantees:
\begin{itemize}
\item that the passwords generated with the help of a pathword are
adequate ({\em i.e.} that they are not easy to guess),
\item that managing pathwords is more user friendly than managing
passwords and that pathwords are less amenable to problems typical of
passwords.
\end{itemize}

\tableofcontents

\section{Passwords: they are useful only if they are robust}

Assume to have a service $S$ which must be accessed only by trusted
users. The authentication of a user $U$ is accomplished by having $U$ to
provide for an identity (usually a {\em user name} $u$) and for a {\em
secret password} $p$. The secret must be known only to $U$, to $S$ and to no
other and it must match the secret that the service already know.

This schema allows $S$ to check that a user $U'$, faking for $U$, is not
the intended user, since $U'$, by definition, is not able to provide to
$S$ the $p$ that is linked to $u$. As usual, we assume that the
identities are publicly known, or that an attacker can easily discover
them, and that all that must be kept secret are indeed the passwords.

Since $p$ is known only to $U$, this system is robust only if the
following assumptions hold:
\begin{enumerate}
\item\label{item1} $p$ is not ``easy'' to guess --- we will make clear the exact
formal meaning of ``easy'' in a moment,
\item $U$ has a simple and friendly way of managing its own many $p$,
corresponding to its many identities in its many services, otherwise $U$
can be tempted to write them down somewhere, or to use ``simple'' $p$
that are amenable of being stolen or guessed.
\end{enumerate}

Of course, the above list is not complete, since in a real-world case
there can be many other issues to be considered, depending on the
technology, on the relevance of the accessed services or on the
particular characteristics of the interaction between $U$ and $S$.

From now on, when we talk about services, we assume them to be like
those mentioned in the introduction (terminal login, access to email or
to web-based applications, and so on), where an actor has to provide its
own data to a computer program, through a keyboard or any other similar
device: this range of services, far from being complete, is of great
relevance to the daily practices of many Internet users.

\subsection{Easy passwords are an hard problem}

What is an ``easy'' password in the sense of item \ref{item1} above ? It
is a secret that is not so difficult to discover.

Assume, without loss of generality, that $p$ is simply a word in the
full language $A^*$ of an alphabet $A$. If we know that $p$ is of length
$n$, then there are $|A|^n$ possible guesses for $p$. This means, that,
if $p$ is randomly chosen, and there are no biases, the probability of
guessing it is only $|A|^{-n}$ and that there need approximately
$|A|^n/2$ attempts to guess it. We associate to each service $S$ a {\em
time frame} $T_S$ that measures how long it takes, on average, to
guess one $p$ for $S$, and we say that that $p$ is {\em adequate} for $S$
when the estimated time to guess is greater that $T_S$. A $p$ which
is not adequate, is {\em easy}.

Of course, in a perfect world, $T_S$ would be infinite, but this
requirement is clearly impossible. For example, if we take $A$ as the
binary alphabet $\{0,1\}$, $T_S$ as one year, and we assume that an
attacker can check by brute force $10^6$ passwords each second, we have
that an adequate $p$ must be strictly more that 46 bits long since:

\[2^{45} > 10^6 \mbox{(pwd/sec)} * 3600 \mbox{(sec/hour)} * 
24  \mbox{(hour/day)} * 365 \mbox{(day/year)} \]

If we assume that the attacker gets stronger or weaker computational
power, the estimate changes consequently.

Forty-six bits of password are not too many: they are guaranteed by a 7
letter random string chosen from the ASCII alphabet (which has $2^7$
characters in it). In practice, we already face a problem, since a
randomly chosen word in $\mbox{ASCII}^7$ can be very hard to remember
and technically very hard to type on an ordinary keyboard, due to the
presence of codes that are interpreted as control characters. If we
stick to a more ordinary alphabet of 16 letters, like the hexadecimal
code, we need a twelve letter password (which really is 48 bits long):
it surely is simpler to write (since we have only ten digits and the
letters A, B, C, D, E and F) but it can be even harder to
remember. Please try to remember this string, you will be asked
about it later: 
\begin{quotation}
{\sl AC43 A172 E1CB 879D}
\end{quotation}

Remembering many different passwords is a heavy burden, recognised by
many researchers. Many security practices (see \cite{NIST-SP-12} and
\cite{NIST-SP-14}, for example) warn against writing down complex or
long passwords, since this helps potential attackers. On the other side,
the same practices advise against using common words or strings that
have a meaning since this makes the passwords amenable to {\em
dictionary attacks}. In both cases it is entirely a user problem to
manage their (many) passwords in order to fulfil security requirements
and to be able to use the same passwords efficiently. In is reported in
\cite{schneier} (pages 104 - 105) that ``\dots entropy (\dots) of
standard English at less than 1.3 bits per characters; passwords have
less than 4 bits of entropy per character'', as opposed to the
theoretical 8 bits of entropy of an ASCII character.

\section{A simple solution: write them down}

Some solutions to the above {\em conundrum} have been proposed, like
having external tokens or devices that provide the {\em secret} that
uniquely identifies the user, or using helper programs that store
securely the passwords.

In the following we will propose an alternate schema, called {\em
pathwords}, which we believe is:
\begin{itemize}
\item secure ({\em i.e.} it allows the management of adequate
passwords),
\item user friendly ({\em i.e.} it is easy to devise helper
applications, to remember and use passwords, and so on).
\end{itemize}

The following example explains the idea behind the {\em pathword}
approach. 

\begin{figure}[h]
\begin{center}
\begin{tabular}{|c|c|c|c|c|c|}
\hline
$a$ & $c$ & $e$ & $2$ & $3$ & $4$ \\
\hline
$a$ & $1$ & $6$ & $f$ & $7$ & $2$ \\
\hline
$d$ & $2$ & $a$ & $1$ & $9$ & $4$ \\
\hline
$f$ & $c$ & $f$ & $a$ & $9$ & $6$ \\
\hline
$e$ & $1$ & $b$ & $5$ & $b$ & $c$ \\
\hline
$8$ & $7$ & $3$ & $4$ & $d$ & $9$ \\
\hline
\end{tabular}
\caption{First example}
\label{example}
\end{center}
\end{figure}

How easier is an attacker job if we state that our own secret password is 
stored in the diagram of picture \ref{example} ? By ``stored'' we mean
that we are able to read it out from the picture, with no other help. 

Quick, now: which was the secret hexadecimal word of the previous
section ? Have you been able to remember it correctly ? If the answer
is yes, congratulations for your memory, otherwise the following hint
may be useful.

\begin{figure}[h]
\begin{center}
\begin{tabular}{|c|c|c|c|c|c|}
\hline
$a^1$ & $c^2$ & $e$ & $2$ & $3^4$ & $4^3$ \\
\hline
$a^5$ & $1^6$ & $6$ & $f$ & $7^8$ & $2^7$ \\
\hline
$d$ & $2$ & $a$ & $1$ & $9$ & $4$ \\
\hline
$f$ & $c$ & $f$ & $a$ & $9$ & $6$ \\
\hline
$e^9$ & $1^{10}$ & $b$ & $5$ & $b^{12}$ & $c^{11}$ \\
\hline
$8^{13}$ & $7^{14}$ & $3$ & $4$ & $d^{16}$ & $9^{15}$ \\
\hline
\end{tabular}
\caption{Annotated example}
\label{example:2}
\end{center}
\end{figure}
 
The numbers appearing at the exponent of some letters indicate the order
to follow to read out the secret word: first the letter $a$, then $c$,
and so on, ending with $9$ and $d$. Another picture shows even more
clearly the pattern followed to read out the password.

\begin{figure}
\begin{center}
\scalebox{0.4}{\includegraphics{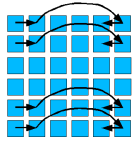}}
\end{center}
\caption{Path example}
\label{example:3}
\end{figure}

It is our position that remembering patterns (or {\em paths}) like those
depicted in picture \ref{example:3} is easier than remembering
passwords, that this practice leads people to use stronger passwords
and that no security is lost in the process. By ``security'', here, we
mean that the practice of remembering passwords with the help of
patterns like those sketched above does not help an attacker
significantly.

\section{Analysing pathwords}

Informally, a {\em pathword} $\pi$ is any walk on the cells of a table
like that previously depicted. Formally, a pathword $\pi$ is a function
between elements of some space $\mathcal{S}$ (in the previous example,
$\mathcal{S}$ is the space of 6x6 matrices over the hexadecimal code
alphabet) to the strings over another alphabet. For the sake of
simplicity, and with no loss of generality, we assume that the two
alphabets are always the same. It is easier to represent pathwords with
diagrams like those of figure \ref{example:3}, but any other
representation is obviously equivalent. For example, the transformation
given by figure \ref{example:3} is 

$$ \pi(\{a_{i,j} | 1 \leq i,j \leq 6\}) = a_{1,1} a_{1,2} a_{1,6}
a_{1,5} a_{2,1} \dots a_{5,5} a_{6,1} a_{6,2} a_{6,6} a_{6,5}$$

Some key points to be considered are the following:
\begin{itemize}
\item users should have more than one pathword, perhaps a least three or 
four, perhaps clustered around different security concerns (for
example, an ``easy'' pathword like the one above to read out password to
access less important services, another one for more important services
and a more complex one to read out password of really sensitive services),
\item the diagrams themselves can even be public, since they convey no
useful information to potential attackers (see about this in section
\ref{robust}),
\item users should not try to make their passwords easier to remember or
to build them following rules: in fact doing this will lower the
effective complexity of the password, making it easier to guess.
\end{itemize}

The key question to be answered is: ``Assuming the adoption of a
pathword, does this undermine the security of the identification process
? If not, under which further assumptions ?'' The remaining of the paper
is devoted to examining this question.

\subsection{Robustness of pathwords vs passwords}\label{robust}

Let's define as $d$ any diagram (for example a table) on an alphabet
$A$. Then a pathword is a function $\pi: d \longrightarrow p$ which maps
$d$ to a $p$ in $A^*$. Given a length $n$, the probability of guessing a
random string in $A^n$ is $|A|^{-n}$. If $d$ contains only a subset $A'$
of letters, and $\pi$ does not have repetitions ({\em i.e.} it never
passes twice on any cell), then the number of possible sequences is
lower-bounded by 

$$ \prod_{j=1}^n (|A'| - (j-1)) $$

Since we have no constrains on $d$, other than the usability of the
proposed solution, we can declare that the process of generating $d$ is
not completely random, and that $A'=A$. The ratio $r$ between the number
of sequences obtained by $\pi$ and the number of sequences of a
perfectly random process is

$$ r = \frac{\prod_{j=1}^n (|A| - (j-1))}{|A|^n}
     = \prod_{j=1}^n (1 - \frac{j-1}{|A|})
    \geq (1 - \frac{n-1}{|A|})^n $$

if we set $k=|A|/(n-1)$ then we have

$$ r  \geq (1 - \frac{1}{k})^{1+|A|/k}$$

if $|A| >> n$, $|A|$ is big and $n$ is small, also $k$ is big (since it
is approximatively a $n^{\mbox{th}}$ fraction of $A$) and $(1 - 1/k)^k$
is approximable by $e^{-1}$, so

$$ r \geq \frac{1}{e^{|A|/k^2}}$$

but $|A|/k^2 = (n-1)/k << 1$. This means that $r$ is not far from 1 (in
the worst case, if $A$ is comparable with  $k^2$, $r$ is still bigger
that $e^{-1}$ or approximately 37\%).

\teo{\label{th:1}
Assume that $A$ is large and $n$ is small, as in the above analysis. 
Given a password $p \in A^n$, and given a diagram $d$ where all the
letters in $A$ appear, it is possible to have a $\pi(d)$ of the
same strength of $p$ just taking $\pi(d) \in A^{n+1}$
}
{ Since in the worst case the ratio between the number of available
sequences and the full set is only of $e^{-1}$, this means that the gain
of a possible intruder is less than $e$ times. Choosing a longer
word, of just two more bits, enlarges the password space by four times,
negating the above speedup. Since $A$ is large, by definition, it is
enough to add one single letter.  }

By theorem \ref{th:1}, using pathwords does not imply to have longer
passwords, in the worst case, just one character longer.

\subsection{An interface to a pathword system}

It is not clear yet which is the better trade-off between the size of
the alphabet $A$, the size of passwords $p$ and the best way to deploy
diagrams $d$.

Assume that $d$ is a square matrix of 100 elements and that the alphabet
is composed of all the couples of digits ({\em i.e.} the alphabet of the
100 ``letters'' ranging from 00, 01, \dots to \dots 98, 99).

Notice that to have a 64 bits password $p$, $p$ must be at least ten
characters long.  A pathword of length ten is not too complex to
remember and can be even something with very few structure in it, and
yet short enough to be remembered. Just to make a comparison, the
pathword of picture \ref{example:3} is structured, the one
\ref{example:4} is less, the one \ref{example:5} has been randomly
generated.

\begin{figure}[h]
\begin{center}
\begin{tabular}{|c|c|c|c|c|c|}
\hline
$\quad$ & $\quad$ & 1 & 4 & $\quad$ & 10 \\
\hline
$\quad$ & $\quad$ & 7 & $\quad$ & $\quad$ & $\quad$ \\
\hline
$\quad$ & $\quad$ & $\quad$ & $\quad$ & $\quad$ & $\quad$ \\
\hline
9 & 3 & $\quad$ & $\quad$ & $\quad$ & $\quad$ \\
\hline
$\quad$ & 6 & $\quad$ & $\quad$ & 2 & $\quad$ \\
\hline
$\quad$ & $\quad$ & $\quad$ & $\quad$ & 5 & 8 \\
\hline
\end{tabular}
\caption{``Triangle-shaped''path}
\label{example:4}
\end{center}
\end{figure}

\begin{figure}[h]
\begin{center}
\begin{tabular}{|c|c|c|c|c|c|}
\hline
$\quad$ & $\quad$ & 1 & $\quad$ & 7 & $\quad$ \\
\hline
$\quad$ & $\quad$ & 3 & 2 & $\quad$ & $\quad$ \\
\hline
$\quad$ & $\quad$ & $\quad$ & $\quad$ & $\quad$ & 8 \\
\hline
6 & $\quad$ & $\quad$ & $\quad$ & $\quad$ & 9 \\
\hline
$\quad$ & $\quad$ & $\quad$ & $\quad$ & 4 & $\quad$ \\
\hline
$\quad$ & 10 & $\quad$ & 5 & $\quad$ & $\quad$ \\
\hline
\end{tabular}
\caption{Random path}
\label{example:5}
\end{center}
\end{figure}

In the above example we have by explicit calculation that $r \approx
63\%$.  This means that it can be a realistic scenario. Notice that a 10
x 10 square table is not very big: it can be easily shown on a web page,
even on a PDA or printed on a sheet of paper of the size of a credit
card.

\section{Discussion}

The following are some question that naturally arise.

{\em What happens if a user loses a pathword? And if it stolen?}

If a user has just one single pathword, and if it is stolen or lost, this
can be a big problem, since an attacker will be able to access all the
users'passwords, just by reading them from the diagrams. If a user had
many pathwords, the loss is mitigated by the reduced number of
accessible passwords.

Comparing the loss of a password to the loss of a pathword, losing the
pathword is potentially worse, since it subsumes the simultaneous loss
of many passwords.

{\em Why should be easier to remember a pathword instead of a password?}

It should be easier for two reasons:
\begin{enumerate}
\item a user should have few pathwords, instead of passwords by the dozen,
\item humans are better suited to remember visual patterns than
random ({\em i.e.} structure-less) strings.
\end{enumerate}

{\em Is it not enough to keep passwords in a secure device --- like a
PDA with some application --- in some encrypted form?}

Maybe. But notice that if a secure solution is available to manage
passwords, there should be no reason, in principle, for not using it
also to manage pathwords. 

Having pathwords, instead of passwords, exposes the user to no further
risk, if the application is secure.

{\em Are passwords really not secure, in practice?}

We are not aware of definitive conclusions on this issue, but some
previous works point out that:
\begin{itemize}
\item {\em real} passwords are not truly random and are exploitable with
brute force attacks,
\item people reuse the same passwords, sometimes even for services of
different importance (for example, exposing the risk to have their
electronic bank account accessed with the same credentials used to
post on a public web forum devoted to their hobbies),
\item if people are forced to use really strong passwords, or to change
them frequently, they tend to write them down somewhere, devising their
own tricks to camouflage them,
\item changing passwords is an issue: people tend to reuse two
passwords, switching between them ({\em i.e.} having a summer password
and a winter password), or to generate a new password appending or
changing some character to their previous password ({\em i.e.}
passwordA, passwordB, and so on). Technical solutions avoid these
behaviours, but empirical evidence shows that changing passwords 
frequently really annoys users.
\end{itemize}

Notice how pathwords address the above issues:
\begin{itemize}
\item passwords read from pathwords tend to be adequate,
\item passwords are never reused: in fact, since a diagram can be
generated by the service to be accessed, there is a practical guarantee
that no two diagrams are equal,
\item the number of pieces of information that has to kept secret is
low, and so there is less pressure to devise ways of remembering them,
\item changing passwords can be an issue no more: a service can even
change the password each time it is accessed, just by providing the user 
with a different diagram each time.
\end{itemize}

\bibliographystyle{plain}

\bibliography{security}

\end{document}